\documentclass[prd,nofootinbib,twocolumn,english]{revtex4}
\usepackage{amsmath}
\usepackage{amsfonts,color}
\usepackage{tensor}
\usepackage{amssymb,float}
\usepackage[utf8]{inputenc}
\usepackage{color}
\usepackage[unicode=true,pdfusetitle,
 bookmarks=true,bookmarksnumbered=true,bookmarksopen=true,bookmarksopenlevel=3,
 breaklinks=false,pdfborder={0 0 1},backref=false,colorlinks=true]
 {hyperref}
\usepackage{enumitem}
\usepackage{graphicx}
\usepackage{appendix}
\usepackage{hyperref}
\usepackage{lipsum}
\usepackage{url}
\usepackage{mathrsfs}
\usepackage{amsbsy}
\usepackage{amstext}
\usepackage{amsthm}

\makeatletter
\long\def\dddddot#1{%
  {\mathop {#1}\limits ^{\vbox to-1.4\ex@ {\kern -\tw@ \ex@ \hbox {\normalfont .....}\vss }}}%
}
\long\def\multidots#1#2{%
  \count@=0
  {{\mathop {#2}\limits ^{\vbox to-1.4\ex@ {\kern -\tw@ \ex@ \hbox {\normalfont %
  \loop%
  \ifnum#1>\count@%
  .%
  \advance\count@ by1%
  \repeat%
  }\vss }}}}%
}
\makeatother

\newcommand{\dd}{\textrm{d}}

\begin{document}

\title{Exact solutions of Cotton Gravity in its Codazzi formulation}

\author{Roberto A. Sussman}
\email{sussman@nucleares.unam.mx}
\affiliation{Instituto de Ciencias Nucleares, Universidad Nacional Aut\'onoma de M\'exico, 
Circuito Exterior C.U., A.P. 70-543, M\'exico D.F. 04510, M\'exico.}

\author{Sebastián N\'ajera}
\email{najera.sebastian@ciencias.unam.mx}
\affiliation{Instituto de Ciencias Nucleares, Universidad Nacional Aut\'onoma de M\'exico, 
Circuito Exterior C.U., A.P. 70-543, M\'exico D.F. 04510, M\'exico.}

\begin{abstract}
The ``Codazzi formulation'', based on a Codazzi tensor,  provides a more robust and straightforward theoretical framework for `Cotton Gravity'' (CG) than its original formulation in terms of the Cotton tensor. Using this formulation we provide a self-consistent procedure to generate  non-trivial exact solutions in CG that generalize well known General Relativity (GR) solutions. We re-derive a known CG solution that generalizes the Schwarzschild solution of GR, showing that it is the unique vacuum solution of static spherical symmetry in CG, extending this result to a CG generalization of the Reissner-Nordstrom solution of GR, all of which places a strong case supporting the fulfillment of Birkhoff's theorem.  When applied to Friedman-Lema\^\i tre-Robertson-Walker (FLRW) models CG naturally identifies the $\Lambda$CDM model as the unique FLRW dust model with constant negative spatial curvature. We also obtain CG generalizations of the Lema\^\i tre-Tolman-Bondi (LTB) and Szekeres dust solutions of GR, allowing for time and space dependent changes from decelerated to accelerated evolution, without necessarily assuming a dark energy source. The CG generalization of  static perfect fluid spheres allows in the weak field regime to model the flattening of rotation velocities in spherical galactic systems without assuming dark matter. We also generalize non-static spherically symmetric perfect fluid solutions with a shear-free 4 velocity.  Our results suggest the need for further research using the Codazzi formulation to explore the potential for applications of CG to current open problems in gravitational systems.
\end{abstract}

\maketitle

\section{Introduction}.-  

Detection of dark sector sources and the understanding of their fundamental properties has remained elusive. This  has motivated the search of gravity theories alternative to General Relativity (GR).  In 2021 Harada \cite{harada2021emergence} proposed a new gravity theory, denoted by ``Cotton Gravity'', which generalizes GR through the rank-3 Cotton tensor \cite{cotton1899varietes,ellis2012relativistic}. The left hand side of the field equations containing the Cotton tensor can be derived from the conformal action \cite{mannheim1989exact}, but varying the connection and keeping the metric fixed. Harada showed that all solutions of GR are also solutions of the new theory, deriving a spherically symmetric vacuum solution that generalizes the Schwarzschild solution of GR. In a follow up article \cite{harada2022cotton} he also showed that in its Newtonian limit the theory allows for fitting galactic rotation curves without assuming the existence of dark matter
\footnote{Harada proposed a separate theory denoted by ''Conformal Killing Gravity'' (CKG) \cite{Harada_2023a,Harada_2023b} (see also \cite{Mantica_2023}), which is a manifestly distinct from ``Cotton Gravity''.}.

Harada's theory in its original formulation (hereafter the ``Cotton formulation'') exhibits a number of shortcomings: it cannot be applied to non-vacuum conformally flat spacetimes  (such as FLRW models) because the Cotton tensor vanishes identically when the Weyl tensor vanishes. Also, it leads to ambiguities in identifying field sources and involves third order field equations, whose integration presents a formidable technical problem even for very symmetric spacetimes. 

CG has been subjected to criticism, but only when formulated in terms of the Cotton tensor (we provide our own critique on the Cotton formulation in Appendix B).  Bargue\~no \cite{bargueno2021comment} argued that CG does not provide a real extension of GR, a claim responded by Harada  \cite{harada2021reply} by emphasizing that non-trivial solutions of the field equations can be derived that are not GR solutions. A more recent critique by Clement and Nouicer \cite{clement2024} claimed that ``Cotton Gravity'' is an ill posed ``not predictive'' theory, plagued by significant ambiguities and inconsistencies. However, as we argued in \cite{response2024},  this critique is incorrect and all its points are strictly applicable only to shortcomings of the Cotton formulation.     

An alternative formulation (denoted hereafter the ``Codazzi formulation'') of ``Cotton Gravity'' was found by Mantica and Molinari \cite{mantica2023codazzi}, based on the condition that defines a Codazzi tensor applied to a second order symmetric tensor ${\cal C}_{ab}$ constructed by a simple algebraic combination of the Einstein (or Ricci) tensors, energy momentum tensors of GR and their traces (the mathematical equivalence of both formulations is shown in Appendix A). We show in this article that the Codazzi formulation provides a more robust and self-consistent theoretical (and practical) framework than the Cotton formulation, avoiding its  known shortcomings, including the points raised in \cite{clement2024}.

In the present article  we use the Codazzi formulation to develop a self-consistent procedure to obtain from known GR solutions their extension as exact CG solutions. The procedure keeps the same metric and energy momentum tensor of the GR solutions, but introduces modifications in their evolution and/or structural equations, achieving dynamical effects absent in GR. We re-derive Harada's extension to GR of the Schwarzschild solution  \cite{harada2021emergence}, showing that it is the unique vacuum solution in static spherical symmetry and finding also the Reissner-Nordstrom CG analogue of a charged point mass.  We also obtain FLRW models whose extension to CG identifies the cosmological constant as constant spatial curvature.  Inhomogeneous CG dust solutions are obtained that generalize  inhomogeneous GR dust solutions:  Lema\^\i tre-Tolman-Bondi (LTB) and quasi-spherical Szekeres dust models, as well as spherically symmetric static perfect fluid spheres and non-static perfect fluid solutions with a shear-free 4-velocity (which \cite{mantica2023codazzi} denote as ``Generalized Stephani'' solutions).  Our procedure can be complemented and made more rigorous with the results by Mantica and Molinari that examine the compatibility between Codazzi tensors and  various assumptions of sources and their connection with known GR solutions \cite{mantica2023codazzi}, as well as  comparing FLRW models in CG and in various alternative gravity theories \cite{mantica2023friedmann}. 

The CG extension of  FLRW, LTB and Szekeres solutions modifies their spatial curvature (Ricci scalar of rest frames), which modifies their Hamiltonian cnstrain (Friedman equation) and introduces modification of the dynamics that allow dynamical degrees of freedom absent in their GR versions. In particular, an accelerated expansion is possible in FLRW and LTB dust models, without  assuming a dark energy source. The specific $\Lambda$CDM model can be identified as the unique FLRW dust model in CG with constant negative spatial curvature. 

In the case of static spheres, we show that in the weak field limit rotation curves in the fluid region of a spherical galaxy can exhibit a flattened profile without assuming dark matter, complementing  the results of Harada in \cite{harada2022cotton} but obtained in the fluid region, not through the Schwarzschild CG vacuum  \cite{harada2021emergence}. We also reproduce the results of Mantica and Molinari in  \cite{mantica2023codazzi} on shear-free solutions with a perfect fluid source.

Evidently, our procedure is a first step to obtain self-consistent CG solutions extending well widely used solutions of GR. It does not exhaust the degrees of freedom afforded by CG and is not the only way to obtain CG solutions. We believe that CG requires further and deeper research, but our procedure  fully complies with a well posed correspondence to GR solutions and corrects misunderstandings that arise in the original Cotton tensor formulation (see \cite{clement2024} and our response in \cite{response2024}). 

The section by section content of this paper is as follows: section II presents CG field equations in the Cotton and Codazzi formulations. while section III outlines our procedure to find non-trivial CG exact solutions. Sections IV, V, VI, VII, VIII and IX respectively present  CG solutions for static spherical symmetry, FLRW models, LTB, Szekeres dust models, spherical symmetric static perfect fluid  solutions and a shear-free 4-velocity. Our conclusions are stated in section.  X. Appendix A shows the equivalence of the Cotton and Codazzi formulations and Appendix B illustrates shortcomings of the Cotton formulation).


\section{From Cotton to Codazzi Gravity}

The field equations in the Cotton formulation of CG originally proposed by Harada \cite{harada2021emergence} (known as ``Cotton Gravity'') are
\begin{equation}
    	C_{abc} = 16\pi {\cal T}_{abc},\label{original}
\end{equation}
where $C_{abc}$ is the Cotton tensor \cite{cotton1899varietes, stephani2009exact,garcia2004cotton} and ${\cal T}_{abc}$ is related to the generalized angular momentum tensor, both given explicitly as
\begin{eqnarray}C_{abc} &=& \nabla_b {\cal R}_{ac}-\nabla_c{\cal R}_{ab}-\frac16\left(g_{ac}\nabla_b{\cal R}-g_{ab}\nabla_c{\cal R}\right),\label{cotton} \\
{\cal T}_{abc} & = & \nabla_{a}T_{bc}-\nabla_{b}T_{ac}-\frac{1}{3}\left(g_{bc}\nabla_{a}T-g_{ac}\nabla_{b}T\right),\label{Tcot}
\end{eqnarray}
with ${\cal R}_{a b}$  and $T_{ab}$ satisfying $\nabla_b T^{ab}=0$ are the Ricci tensor and the energy momentum tensor of GR. 

An alternative  ``Codazzi formulation'' of CG was derived by Mantica and Molinari \cite{mantica2023codazzi} in terms of the condition that defines a Codazzi tensor
\begin{equation}\nabla_b{\cal C}_{ac} - \nabla_c{\cal C}_{ab} =0,\label{codazzi}\end{equation}
where  ${\cal C}_{ab}$ is the following symmetric tensor  constructed with the Einstein tensor $G_{ab}$,\,\,$T^{ab}$ and $g_{ab}$ 
\begin{equation} {\cal C}_{ab}=G_{ab}- 8\pi T_{ab} -\frac13 (G-8\pi T)\,g_{ab},\label{precodazzi}
\end{equation}
where  $G=g^{cd}G_{cd},\,\,  T=g^{cd}T_{cd}$ and we assume that $\nabla_b T^{ab}=0$.  

While both formulations: \eqref{original}-\eqref{Tcot} and  \eqref{codazzi}-\eqref{precodazzi} are mathematically transformable into each other  (see proof in Appendix A), there are significant theoretical and practical differences in how they work.  As mentioned in the introduction, the Cotton formulation of CG exhibits several problematic features that do not occur (or can be better addressed) in the Codazzi formulation. Specifically, we show using the Codazzi formulation of CG: 
\begin{itemize}
\item  Can be applied to non-vacuum conformally flat spacetimes because it does not require computing beforehand the Cotton tensor, which vanishes identically if the Weyl tensor vanishes.
\item Allows to identify unambiguously the field sources. The Codazzi formulation does not involve the generalized angular momentum tensor ${\cal T}_{abc}$ appearing in the right hand side of \eqref{original}, which is not a generic matter-energy source and introduces significant ambiguity. It depends on $T_{ab}$ but also on specific spacetimes by involving covariant derivatives of  $T_{ab}$. While $T_{ab}=0$ implies ${\cal T}_{abc}=0$, the converse statement is in general false, hence ${\cal T}_{abc}=0$ does not necessarily lead to vacuum sollutions. 
\item Involves directly the Einstein tensor and $T_{ab}$ in \eqref{precodazzi}, which facilitates identifying the correspondence to a GR limit, an important consistency requirement of  alternative gravity theories.  
\item Its field equations are much more tractable than the field equations \eqref{original}, a complicated system of non-linear third order equations that make it difficult to prescribe initial conditions and control the differentiation process at first and second orders. The Codazzi formulation splits this third order process in two stages: first a second order problem to obtain ${\cal C}_{ab}\ne 0$   in \eqref{precodazzi} and then the extra differentiation in \eqref{codazzi}.  This splitting in lower orders would be extremely difficult (or impossible) if proceeding with the third order the Cotton formulation. 
\end{itemize}

The Codazzi formulation allows two options for the energy momentum tensor:
\begin{itemize}
\item Use the same $T_{ab}$ of GR, satisfying $\nabla_b T^{ab}=0$,  in \eqref{precodazzi} guided by the Correspondence Principle with GR. 
\item An ``effective'' energy momentum tensor.  For any CG solution with ${\cal C}_{ab}\ne 0$ satisfying  \eqref{codazzi}, it is possible to express ${\cal C}_{ab}$ in \eqref{precodazzi} as ``effective'' Einstein's equations :
\begin{equation} G_{ab} = 8\pi [T_{ab}+T_{ab}^{(\hbox{\tiny{eff}})}],\qquad  T_{ab}^{(\hbox{\tiny{eff}})} = {\cal C}_{ab}-{\cal C}\,g_{ab},\label{effective}
\end{equation}
where ${\cal C}=g^{cd}{\cal C}_{cd}$. 
\end{itemize}
We remark that expressing field equations in the form \eqref{effective} does not imply that CG is equivalent to GR. It just a parametrization of the field equations whose usage is a common practice in alternative gravity theories (see examples in $f({\cal R})$ theories  \cite{odintsov2019effects,odintsov2018dynamical}). It  involves an ``effective'' energy momentum tensor  $T_{ab}^{(\hbox{\tiny{eff}})}$ made of geometric terms that emerge from the alternative theory ``passed'' to the right hand side facing $G_{ab}$. Whether it is useful or not depends on the possibility to assign physical (or at least plausible) properties  to these geometric terms. As shown by Mantica and Molinari \cite{mantica2023friedmann}, CG equations in the form \eqref{effective} provide interesting description of dark fluids and allow to compare CG with other alternative theories. 

In what follows we use the Codazzi formulation to obtain various exact CG solutions  that generalize and extend known GR solutions. We will proceed keeping as reference the correspondence with the GR solutions. Unless stated otherwise, we will consider only the same $T_{ab}$ as in GR (not the effective tensor \eqref{effective}).   

\section{From General Relativity to Cotton Gravity}

It is useful to rewrite the field equations in the Codazzi formulation \eqref{codazzi}-\eqref{precodazzi} in a more compact form 
\begin{eqnarray}{\cal C}_{ab} &=& {\cal S}_{ab}-8\pi \mathscr{T}_{ab} \qquad   \hbox{ is a Codazzi tensor }\nonumber\\
 &\Rightarrow&\quad  \nabla_aC_{bc}-\nabla_bC_{ac}=0,\label{codazzi2}\end{eqnarray}
where ${\cal S}_{ab}$, the Schouten tensor, and $\mathscr{T}_{ab}$ are
\begin{eqnarray} {\cal S}_{ab} &=&  {\cal R}_{ab}-\frac16{\cal R} g_{ab}= G_{ab}-\frac13 G g_{ab},\label{schouten}\\
\mathscr{T}_{ab} &=& T_{ab}-\frac13 T g_{ab}.\label{Tgorda}
\end{eqnarray}
It is illustrative to rewrite ${\cal C}_{ab}$ in \eqref{codazzi2} as 
\begin{equation}{\cal C}_{ab}=G_{ab}-8\pi T_{ab}-\frac13\left(G-8\pi T\right)g_{ab},\label{precodazzi2}\end{equation}
clearly showing that for any solution of GR: $G_{ab}-8\pi T_{ab}=0$ and $G-8\pi T= g^{cd}(G_{cd}-\kappa T_{cd})=0$, leading to :
\begin{equation}{\cal C}_{ab}={\cal S}_{ab}-8\pi \mathscr{T}_{ab}=0,\quad \hbox{GR solution}\label{codazziGR}\end{equation}
so that all GR solutions are trivial solutions of CG in \eqref{codazzi2}. Therefore, non-trivial CG solutions that generalize GR solutions follow from considering \eqref{codazziGR} and rewriting  \eqref{codazzi2} as 
\begin{eqnarray}{\cal C}_{ab} &=& {\cal S}_{ab}-8\pi \mathscr{T}_{ab} =  \mathscr{C}_{ab} \ne 0,\label{codazzi3a}\\
 &\Rightarrow&\quad  \nabla_a\mathscr{C}_{bc}-\nabla_b\mathscr{C}_{ac}=0,\label{codazzi3b}\end{eqnarray}
where the tensor $\mathscr{C}_{ab}$ must be tested to be Codazzi tensor by complying with \eqref{codazzi3b}. This tensor can be determined for each spacetime and choice of $T_{ab}$ and/or from the properties of a second rank Codazzi tensors in terms of covariant 4-vectors and tensors  \cite{mantica2023codazzi}. However, in the present article we will use \eqref{codazzi3a}-\eqref{codazzi3b} to obtain $\mathscr{C}_{ab}$ in a case by case basis to generalize GR solutions along the following lines
\begin{enumerate}
\item Compute ${\cal C}_{ab}$ in \eqref{codazzi2} for a given known solution of GR 
\item Substitute Einstein equations in the computed ${\cal C}_{ab}$, which must lead to \eqref{codazziGR}
\item Modify the GR solution without changing the metric and $T_{ab}$ and compute again ${\cal C}_{ab}$ in \eqref{codazzi2}. This will identify $\mathscr{C}_{ab}$ in \eqref{codazzi3a},
\item Compute \eqref{codazzi3b} to obtain the conditions for ${\cal C}_{ab}$ to be a Codazzi tensor, the CG solution follows from solving the resulting constraints,
\end{enumerate}
This procedure clearly distinguishes among non-trivial CG solutions ($\mathscr{C}_{ab}\ne 0$) between vacuum and non-vacuum solutions:
\begin{itemize}
\item Non-vacuum: $\mathscr{T}_{ab}=T_{ab}-\frac13 T g_{ab}\ne 0$, hence:
\begin{equation} {\cal S}_{ab}=\mathscr{C}_{ab}+8\pi\,\mathscr{T}_{ab}\quad {\cal S}_{ab}\ne \mathscr{C}_{ab},\label{nonvac}\end{equation}
\item Vacuum: $\mathscr{T}_{ab}=T_{ab}-\frac13 T g_{ab}= 0$, hence:
\begin{equation}  {\cal S}_{ab}=\mathscr{C}_{ab}\ne 0,\label{vac}\end{equation}
\end{itemize}
Notice that a univocal distinction between vacuum and non-vacuum CG solutions is a much more complicated problem in the Cotton formulation \eqref{original}, since the condition ${\cal T}_{abc}=0$ does not imply $T_{ab}=0$ and can be satisfied by nonzero energy momentum tensors.   

\section{Spherically symmetric static vacuum solutions}   

The first solution found by Harada is the CG analogue of the Schwarzschild solution \cite{harada2021emergence}.  We show that the Codazzi formulation shows this to be the unique CG vacuum solution in spherical static geometry, further extending this solution  to a Reissner Nordstrom analogue. 

Consider  the general spherically symmetric static metric
\begin{equation}
ds^{2}=-A(r)dt^{2}+\frac{dr^{2}}{B(r)}+r^{2}(d\theta^{2}+\sin^{2}\theta d\phi^{2}),\label{staticSS}
\end{equation}
whose GR solution is the Schwarzschild solution: $A=B = 1-2M_s/r$, as the  unique solution of  ${\cal S}_{ab}=0$, leading to a trivial solution of \eqref{codazzi2} complying with \eqref{codazziGR}.

For a non-trivial CG solution we need to pass from \eqref{codazziGR} to \eqref{codazzi3a} to obtain $\mathscr{C}_{ab}\ne 0$. To test the vacuum condition $\mathscr{T}_{ab}=T_{ab}-\frac13 T g_{ab}= 0$, we consider the most general energy momentum tensor compatible with \eqref{staticSS} in a comoving frame $u^a=A^{-1/2}\delta^a_t$ 
\begin{equation}
T_{ab} = \rho(r) u_{a}u_{b}+p(r)h_{ab}+\Pi_{ab},\label{TabST}
\end{equation}
where $\Pi_{ab}$ is the anisotropic pressure whose nonzero components are $\Pi_{r}^{r}=-2P,\,\,\Pi_{\theta}^{\theta}=\Pi_{\phi}^{\phi}=P,\,\,\Pi_{t}^{t}=0$, with $P=P(r)$.  It is straightforward to show that inserting \eqref{TabST} in  $\mathscr{T}_{ab}=T_{ab}-\frac13 T g_{ab}= 0$ leads to a unique solution $\rho=p=P=0$, a true vacuum state. Hence,  equation \eqref{codazzi3a} takes the form \eqref{vac}. As a contrast with straightforward determination of the conditions for a vacuum solution in static spherical symmetry, the conditions to identify vacuum solutions in the Cotton formulation is plagued by ambiguities. We illustrate this point in Appendix B.

To find $ \mathscr{C}_{ab} \ne 0$ we remark that in GR $A=B$ is a necessary (not sufficient) condition to obtain  the Schwarzschild solution corresponding to $ \mathscr{C}_{ab} = 0$. This clearly points out finding  $ \mathscr{C}_{ab} \ne 0$ by modifying the Schwarzschild metric functions in a way compatible with a well defined GR limit that satisfies \eqref{vac} and the necessary condition $A=B$. The most general metric modifying Schwarzschild is then
\begin{equation} ds^2 = -\Phi(r) dt^2 + \frac{dr^2}{\Phi(r)} + r^2(d\theta^2+\sin^2\theta d\phi^2)\label{Schw1} \end{equation}
where $\Phi(r)$ is a completely arbitrary smooth function. This leads to  ${\cal S}_{ab}=\mathscr{C}_{ab}\ne 0$ with nonzero components 
\begin{eqnarray}\mathscr{C}^\theta_\theta =\mathscr{C}^\phi_\phi=-2 \mathscr{C}^t_t =-2\mathscr{C}^r_r 
= \frac{\Phi_{,rr}r^2-2\Phi_{,r}r-4\Phi+1}{6r^2},\nonumber\\ \label{CabVac}\end{eqnarray}
which inserted in  \eqref{codazzi3b}  leads to the following linear third order differential equation 
\begin{equation}   r^3 \Phi_{,rrr} +r^2 \Phi_{,rr}-2r \Phi_{,r}+2\Phi-2=0,\label{ODE3}\end{equation}
whose solution is
\begin{equation}
\Phi(r) = 1 - \frac{2M_s}{r} +\gamma\,r + \frac{8\pi}{3}\,\Lambda\,r^2. \label{Schw2}
\end{equation}
where we have identified the integration constants as the Schwarzschild mass $2M_s$ and  the cosmological constant (emerging naturally as an integration constant), while the linear term $\gamma r$ is inherent of CG. This is the solution found by Harada in  \cite{harada2021emergence}.  

This solution is unique, since we emphasize that $A=B$ is a necessary (not sufficient) condition for \eqref{vac} for the metric \eqref{staticSS}. Any solution with $A\ne B$ will produce ${\cal C}_{ab}={\cal S}_{ab}-8\pi \mathscr{T}_{ab}=0$ in GR, but with  ${\cal S}_{ab}=8\pi \mathscr{T}_{ab}\ne0$, while in CG it will lead to \eqref{nonvac}, not to the vacuum equations \eqref{vac}. so it cannot be a CG vacuum solution of \eqref{staticSS}. This is an important result, suggesting that CG complies with Birkhoff's theorem with  \eqref{Schw2} being the unique vacuum solution for static spherical symmetry in CG, though a rigorous proof is outside the scope of this paper. 

This result does not arise with the Cotton formulation (see Appendix B), since $C_{abc}=0$ in \eqref{original} leads to a complicated third order differential equation that couples the metric functions $A,\,B$ and their derivatives, so it admits an infinite number of solutions. Setting $A=B$ when computing \eqref{original} for \eqref{staticSS} leads to \eqref{ODE3}-\eqref{Schw2}, but there is an infinite number of solutions for $A\ne B$ that comply with ${\cal T}_{abc}=0$ and thus $C_{abc}=0$, which are erroneously regarded as vacuum solutions, when they are strictly solutions with vanishing generalized angular momentum (a confusion that is clearly identified in the critique of \cite{clement2024}). In fact, as shown in Appendix B, substitution of \eqref{TabST}, the most general energy momentum for \eqref{staticSS}, into condition ${\cal T}_{abc}=0$, with ${\cal T}_{abc}$ given by \eqref{Tcot}, leads to a fluid with nonzero $\rho,\,p,\,P$ (see \cite{response2024}), further showing that  these are not vacuum solutions of CG. 

In a recent article Gogberashvili and Girgviliani  \cite{gogberashvili2023general} explored solutions of $C_{abc}=0$ with $A$ taking the form \eqref{Schw2} and $B\ne A$ kept as free function. They analyzed their properties, but none of them is a vacuum solution and none bears any relation with a point mass solution, though this can only be seen through the Codazzi formulation. 

It is straightforward to extend the CG vacuum solution \eqref{Schw2} to a Reissner-Nordstrom CG solution with the energy momentum tensor $T^a_b = -Q/r^4\times \hbox{diag}[-1,1,1-1]$ of a point mass with  electric charge $Q$. This introduces the following modifications of \eqref{codazzi3a} and \eqref{ODE3}
\begin{eqnarray}\mathscr{C}^\theta_\theta &=&\mathscr{C}^\phi_\phi=-2 \mathscr{C}^t_t =-2\mathscr{C}^r_r\nonumber\\
&=& \frac{\Phi_{,rr}r^2-2\Phi_{,r}r-4\Phi+1}{6r^2}-\frac{8\pi Q}{r^4},\label{Crn} \end{eqnarray}
\begin{equation}
   r^3 \Phi_{,rrr} +r^2 \Phi_{,rr}-2r \Phi_{,r}+2\Phi-2+\frac{96\pi Q}{r^2}=0,\label{ODErn}\end{equation}
leading to
\begin{equation}
\Phi(r) = 1 - \frac{2M_s}{r} +\frac{8\pi Q}{r^2}+\gamma\,r + \frac{8\pi}{3}\,\Lambda\,r^2. \label{RN}
\end{equation}
which is the CG analogue of the Reissner-Nordstrom solution, recovering \eqref{Schw2} with $Q=0$.   

\section{FLRW solutions in Cotton Gravity}

FLRW models are non-vacuum conformally flat spacetimes, hence it is impossible apply to them the Cotton formulation in \eqref{original}, since the Cotton tensor vanishes identically and the field equations \eqref{original} become inconsistent. However, as shown in \cite{mantica2023friedmann}, these spacetimes can be handled by the Codazzi formulation \eqref{codazzi}-\eqref{precodazzi}. Consider the general spacetime compatible with the Robertson-Walker metric in spherical coordinates
\begin{eqnarray} ds^2 &=& -dt^2+a^2(t) \left[\frac{dr^2}{1-kr^2}+r^2(d\theta^2+\sin^2\theta d\phi^2) \right],\nonumber\\ \label{flrw}\\
T^{ab}&=&(\rho+p) u^au^b+pg^{ab},\qquad u^a=\delta^a_t,\label{flrw2}
\end{eqnarray} 
with $k=k_0 H_0^2|\Omega_0^{(k)}|,\,\,k_0=0,\pm\,1$. For an equation of state relating $p$ and $\rho$,  GR solutions follow from solving 
\begin{eqnarray} \nabla_bT^{ab}&=& 0 \quad \Rightarrow\quad \dot\rho+\frac{3\dot a}{a}(\rho+p)=0,\label{conseq}\\ 
\dot a^2 &=&  \frac{8\pi}{3}\rho(a)\,a^2-k,\label{friedeq2a}\end{eqnarray} 
The components of ${\cal C}_{ab}$ in \eqref{codazzi2} are  
\begin{eqnarray}{\cal C}^r_r &=& {\cal C}^\theta_\theta =  {\cal C}^\phi_\phi = -\frac{8\pi}{3}\rho +\frac{\dot a^2 +k}{a^2},\label{codRW1}\\
 {\cal C}^t_t &=& 8\pi\left(\frac23\rho+p \right) +\frac{2a \ddot a-\dot a^2-k}{a^2},\label{codRW2} 
\end{eqnarray}
which become from substitution of the conservation and Friedman equations \eqref{conseq}-\eqref{friedeq2a} yields \eqref{codazziGR} ${\cal C}_{ab}=\mathscr{C}_{ab}=0$.

In order to obtain  $\mathscr{C}_{ab}\ne 0$ in  \eqref{codazzi3a}, we need to modify the FLRW models of GR. Since we are keeping the same $T_{ab}$, the only option is to modify the Friedman equation \eqref{friedeq2a} as
\begin{equation} \dot a^2 = \frac{8\pi}{3}\rho(a)\,a^2-k-\gamma{\cal K}(t),\label{friedeqCT}\end{equation}
where ${\cal K}(t)$ is (for the time being) a dimensionless free function and  $\gamma$ is constant with inverse squared length units to be determined further ahead. The components of  $\mathscr{C}_{ab}\ne 0$ in  \eqref{codazzi3a} are
\begin{equation} \mathscr{C}^r_r = \mathscr{C}^\theta_\theta =  \mathscr{C}^\phi_\phi = \frac{\gamma{\cal K}}{a^2},\quad \mathscr{C}^t_t=\frac{\gamma}{a}\left(\frac{\dot{\cal K}}{2\dot a}-\frac{{\cal K}}{a}\right),\label{Cflrw}\end{equation}
which yield \eqref{conseq} when inserted in \eqref{codazzi3b}, so this equation is satisfied.    

Another approach to FLRW models in CG is to modify the sources considering an ``effective'' $T_{ab}^{(\hbox{\tiny{eff}})}$, as in \eqref{effective}, leading also to modified Friedman equations and interesting inferences of the effective fluids with dark sources and with other alternative gravity theories \cite{mantica2023friedmann}.   

However, if $T_{ab}$ is not modified as in  \eqref{effective}, then  ${\cal K}(t)$ is necessarily a geometric degree of freedom brought by CG that modifies the spatial curvature of the models. The modified Friedman equation \eqref{friedeqCT}  changes the expansion scalar $\Theta=\nabla_au^a=3\dot a/a$, which in turn implies a modification of the extrinsic curvature of the hypersurfaces of constant $t$ orthonormal to the 4-velocity: $K_{ab}=h_a^ch_b^d \nabla_bu_a=(\Theta/3)h_{ab}$ that determines the embedding of these submanifolds in the enveloping spacetime. The connection with spatial curvature follows readily from the Gauss-Codazzi equations \cite{thorne2000gravitation,poisson2004relativist,ellis2012relativistic}, leading to
\begin{eqnarray}{}^3{\cal R}&=&16\pi  T_{ab}u^au^b+ K_{ab}K^{ab}-K^2,\qquad K=K^a_a,\nonumber\\
&=& 16\pi\rho-\frac23\Theta^2 = \frac{6[k+\gamma{\cal K}]}{a^2}.\label{KCG2} \end{eqnarray}
where ${}^3{\cal R}$ is the Ricci scalar of the hypersurfaces orthogonal to $u^a$. 

The modification of  ${}^3{\cal R}$ changes the dynamics of the models, as can be seen in the following system of 3 evolution equations and the Friedman equation \eqref{friedeqCT} as the Hamiltonian constraint:
\begin{eqnarray}
\dot\rho &=& -(\rho+p)\Theta,\label{CGa}\\
\dot\Theta &=& -\frac13\Theta^2-4\pi(\rho+3p)-\frac{3\lambda}{2a}\frac{\dd {\cal K}(a)}{\dd a},\label{CGb}\\
\dot a &=& \frac13\Theta a,\label{CGc}\\
\frac{\Theta^2}{9} &=& \frac{8\pi}{3}\rho-\frac{k+\lambda {\cal K}}{a^2},\label{CGd}
\end{eqnarray} 
where $d {\cal K}(a)/da =\dot {\cal K}/\dot a$. The integration of \eqref{CGa}-\eqref{CGd} requires a functional form ${\cal K}={\cal K}(a)$ to be specified, but there is no evolution equation or conservation law for this term. Hence, for a general ${\cal K}(a)$ the system \eqref{CGa}-\eqref{CGd} is incompatible with a well posed initial value formulation that requires finding a unique solution given initial data at an initial hypersurface $t=t_0$, which  only determines  ${\cal K}_0={\cal K}(a(t_0))={\cal K}(1)$, but without an evolution equation that determines ${\cal K}$ for $t\ne t_0$.  While ${\cal K}(a)$ can be prescribed beforehand and could lead to interesting dynamics unachievable by FLRW models in GR, this involves a priori determination of future (even asymptotic) evolution disconnected from initial conditions.

However, there are   two particular cases of FLRW models in CG in which \eqref{CGa}-\eqref{CGd} are compatible with a well posed initial value formulation:
%
%
\begin{eqnarray}{\cal K} &=& \hbox{constant},\qquad  H^2 = \frac{8\pi}{3} \rho -\frac{k+\lambda}{a^2},\label{mod1}\\
%
%
%
%
  {\cal K}&=&a^2,\qquad\qquad H^2 = \frac{8\pi}{3} \rho -\frac{k}{a^2}-\lambda,\label{mod2}\end{eqnarray}
%
%
Models \eqref{mod1} are FLRW models of GR with $\Lambda=0$ and $k$ rescaled, trivial solutions of CG with ${\cal C}^a_b=0$ in \eqref{codRW1}-\eqref{codRW2}.  Models \eqref{mod2} are non-trivial CG solutions with   $\mathscr{C}_{ab}\ne 0$ in  \eqref{codazzi3a},  with components $\mathscr{C}^t_t=0,\,\,\,\mathscr{C}^r_r=\mathscr{C}^\theta_\theta=\mathscr{C}^\phi_\phi=\lambda\ne 0$ in  \eqref{Cflrw}. They are  characterized by constant spatial curvature: ${}^3{\cal R}=6\lambda$, whose sign is determined by the sign of $\lambda$. 

The models \eqref{mod2} above operationally coincide with FLRW models of GR with nonzero $\Lambda$ by identifying $\lambda=(8\pi/3)\Lambda$.  However, there is an important conceptual difference. In GR, $\Lambda$ has units of energy density, so theoretically it is a source (likely an empiric approximation of dark energy), hence it is normalized by $8\pi G/c^4$ to put it in 1/$\hbox{length}^2$ units. In CG $\lambda$ has the same dynamical effect as $(8\pi/3)\Lambda$, but it is a constant spatial curvature, not a source.  Hence, to emphasize the conceptual difference it is better to use a different symbol $\lambda$ and since it already has  1/$\hbox{length}^2$ units, it can be fixed simply as a value proportional to $H_0^2$ that provides a best fit for observations.  

For cosmological applications we assume  $\Lambda>0$ (hence $\lambda<0$) and $\rho=\rho_0/a^3$ to represent cold dark matter (CDM) density as dust (neglecting other constituents like baryons, electrons, photons and neutrinos). We consider  then the models \eqref{mod2} with $\lambda<0$ and assigning its numerical value as $\lambda = -(8\pi/3)\Lambda$ determined by observations. 

%
%
 In particular,  the $\Lambda$CDM is equivalent to the FLRW model (2) with $k =0$, which is the unique FLRW dust model in CG with constant negative spatial curvature. This is an appealing theoretical characterization of the $\Lambda$ in the $\Lambda$CDM model, preferable to that of a rough uncertain late time approximation to an elusive  dark energy source or an empirical construct to fit observations.   Thus, the role of the $\Lambda$CDM model in providing theoretical interpretation to large scale observations associated to an FLRW background can be preserved in CG, with the accelerated expansion driven by its constant spatial curvature, without the need to invoke a dark energy source.

 We can also identify as particular cases of models (2)  the constant curvature vacuum cases setting $\rho=0$, lading to de Sitter ($\lambda = -(8\pi/3)\Lambda$), anti-de Sitter ($\lambda = (8\pi/3)\Lambda$) and Minkwski ($\lambda=0$).

\section{LTB dust solutions in Cotton Gravity}

The spherically symmetric Lema\^\i tre-Tolman-Bondi (LTB) dust solutions of GR provide a simple inhomogeneous and anisotropic generalization of FLRW dust models.  As we show in this section,  non-trivial dust solutions of CG can be obtained based on the LTB metric 
\begin{equation}ds^2 = -dt^2 +\frac{R'^2}{1-K}\,dr^2+R^2 (d\theta^2+\sin^2\theta d\phi^2),\label{LTBmetric}\end{equation}
where  $R=R(t,r),\,K= K(r)$ and a prime denotes $\partial_r$.  For a dust source $T^{ab}=\rho u^a u^b$ in a comoving frame $u^a=\delta^a_t$, the field equations of GR reduce to  
\begin{eqnarray}
\dot R^2 &=& \frac{2M}{R}-K,\label{Flike}\\
8\pi \rho &=& \frac{2M'}{R^2 R'},\label{LTBdust}\\
\nabla_b T^{ab}&=& 0 \quad\Rightarrow\quad \dot \rho+\left(\frac{2\dot R}{R}+\frac{\dot R'}{R'}\right)\rho=0, \label{conseqLTB}
\end{eqnarray}
where $M=M(r)$. 

The components of ${\cal C}_{ab}$ in \eqref{codazzi2} for the GR solution  \eqref{LTBmetric}-\eqref{LTBdust} are 
\begin{eqnarray}
{\cal C}^t_t&=&\frac{16\pi \rho}{3}+\frac{F_0}{3R^{2} R'},\quad {\cal C}^r_r=-\frac{8\pi \rho}{3}+\frac{F_1}{3R^{2}R'},\label{CLTB1}\\
{\cal C}^\theta_\theta &=& {\cal C}^\phi_\phi = -\frac{8\pi \rho}{3}-\frac{F_2}{6R^{2}R'},\label{CLTB2}\\
F_0 &=& 2 R^{2} \ddot R'+4 R R' \ddot R-2 R \dot R \dot R'-R' \dot R^{2}-K R'-R K'\nonumber\\
F_1 &=& 2 R^{2} \dot R'-2 R R' \ddot R+4 R \dot R \dot R'-R' \dot R^{2}-K R'+2 R K',\nonumber\\ 
F_2 &=& 2 R^{2} \ddot R'-2 R R' \ddot R-2 R \dot R \dot R'-4 R' \dot R^{2}-4 K R'-R K'.\nonumber
\end{eqnarray}
substitution of \eqref{Flike}-\eqref{conseqLTB} in \eqref{CLTB1} and \eqref{CLTB2} yields ${\cal C}_{ab}=0$ fulfilling \eqref{codazziGR}. 
To pass from \eqref{codazziGR} to \eqref{codazzi3a} we need to obtain a non-trivial dust CG solution with $\mathscr{C}_{ab}\ne 0$ keeping the  metric \eqref{LTBmetric} and the dust energy momentum tensor. As with FLRW, the only option is to modify the Friedman-like equation \eqref{Flike} as
\begin{equation} \dot R^2 = \frac{2M}{R}-K-{\cal F}(R),\label{Flike2} \end{equation}
where ${\cal F}(R)$ is a dimensionless free function to be determined by \eqref{codazzi3b}. Notice that ${\cal F}={\cal F}(R)$ is the most general form to allow for the  the integration of \eqref{Flike2}. From \eqref{Flike2} we get the nonzero components of $\mathscr{C}_{ab}$
\begin{eqnarray}  \mathscr{C}^\theta_\theta &=& \mathscr{C}^\phi_\phi=-\frac16\frac{d^2{\cal F}}{dR^2}+\frac{1}{3R}\frac{d{\cal F}}{dR}+\frac{2{\cal F}}{3R} ,\label{codLTB1} \\
\mathscr{C}^r_r &=& \mathscr{C}^t_t  = \frac13\frac{d^2{\cal F}}{dR^2}+\frac{1}{3R}\frac{d{\cal F}}{dR}-\frac{{\cal F}}{3R^2}, \label{codLTB2}
\end{eqnarray}
which inserted in \eqref{codazzi3b} yields the following constraint
\begin{equation}\frac{d^3{\cal K}}{dR^3}+\frac{1}{R}\frac{d^2{\cal K}}{dR^2}-\frac{2}{R^2}\frac{d{\cal K}}{dR}+\frac{2{\cal K}}{R^3}=0,\label{consLTB}\end{equation}
whose solution is very similar to the CG Schwarzschild-like solution \eqref{Schw2}
\begin{equation}  {\cal F}(R) = \gamma(r) R+ \frac{\beta(r)}{R}+\lambda(r) R^2,\label{eqPhiLTB}\end{equation}
where we can identify $\beta(r)=2M(r)$ as the effective quasi local mass of comoving dust layers, while $\gamma(r)$ and $\lambda(r)$ are arbitrary functions with inverse length and inverse square length units that appear as integration constants.  Equation \eqref{Flike2} becomes now 
\begin{equation} \frac{\dot R^2}{R^2} = \frac{2M}{R^3}-\frac{K+\gamma R+\lambda R^2}{R^2},\label{Flike3} \end{equation}
Notice that the quadratic term  leads to an $r$-dependent cosmological constant $\lambda(r)R^2$, which need not be imposed, but can be incorporated directly in the modification of the Friedman-like equation \eqref{Flike2} by the particular case $\lambda(r)=(8\pi/3)\Lambda$. The term $\gamma(r) R$ is analogous to the extra linear term in the Schwarzschild CG solution \eqref{Schw2}. 

We emphasize that we have derived a non-trivial CG solution for the LTB metric \eqref{LTBmetric} with a dust source given by \eqref{LTBdust}. While LTB models in GR with a nonzero $\Lambda$ (LTB de Sitter) are also characterized by \eqref{LTBmetric} and \eqref{LTBdust}, the metric functions $R,\,\,R^\prime$ must be now computed with the solution $R(t,r)$ obtained with the CG Friedman-like equation \eqref{Flike3}, which contains the term $\gamma(r) R$ and $\lambda(r) R^2$ thus are necessarily different from those obtained in GR from  \eqref{Flike} (with $\Lambda$ zero or nonzero). 

As with FLRW models, the modification of the Friedman-like equation leading to \eqref{Flike3} also involves a change of the spatial curvature. To illustrate this fact, we stress first that a modified Friedman-like equation modifies the kinematic parameters (expansion scalar $\Theta$ and shear tensor $\sigma_{ab}$) in the extrinsic curvature of the hypersurfaces orthogonal to $u^a$
\begin{eqnarray} K_{ab}&=&  h_a^c h_b^d \nabla_b u_a = \frac{\Theta}{3} h_{ab}+\sigma_{ab},\nonumber\\
K^r_r &=& \frac{\dot R^\prime}{R^\prime},\qquad K^\theta_\theta = K^\phi_\phi =\frac{\dot R}{R}\label{LTBK}\\
\Theta &=& \nabla_au^a = \frac{2\dot R}{R}+\frac{\dot R^\prime}{R^\prime},\label{LTBexp}\\
\sigma^a_b &=& \hbox{diag}[0,-2\Sigma,\Sigma,\Sigma],\quad \Sigma = -\frac13\left(\frac{\dot R^\prime}{R^\prime}-\frac{\dot R}{R}\right),\nonumber\\
\label{LTBshear}
\end{eqnarray} 
Substitution of  \eqref{Flike3} into \eqref{LTBK}-\eqref{LTBshear} and using the Gauss-Codazzi equations \eqref{KCG2} and the LTB dust density \eqref{LTBdust} we get the Ricci scalar of the hypersurfaces orthogonal to $u^a$
\begin{eqnarray} {}^3{\cal R}=16\pi\rho-\frac23\Theta^2+\sigma_{ab}\sigma^{ab}=\frac{2\left[(K+\gamma R+\lambda R^2)\,R\right]^\prime}{R^2\,R^\prime},\nonumber\\\label{3RLTB}\end{eqnarray}
which contains the term $K+\gamma R+\lambda R^2$ in the modified Friedman-like equation \eqref{Flike3}. 

The spatial curvature term that appears in the right hand side of \eqref{Flike3} is the quasilocal scalar (``q-scalar'') associated to ${}^3{\cal R}$ by the average distribution \cite{sussman2013weighed1,sussman2013weighed2}
\begin{equation} {}^3{\cal R}_q =\frac{\int_0^r{{}^3{\cal R}\,R^2\,R^\prime d\bar r}}{\int_0^r{R^2\,R^\prime d\bar r}} = \frac{K+\gamma R+\lambda R^2}{R^2},\label{3Rq}\end{equation}
which applied to the density $8\pi\rho R^2 R^\prime = 2M^\prime$ and the expansion scalar $\Theta\, R^2 R^\prime = \partial_t (R^2 R^\prime)$, leads to 
\begin{equation} \frac{4\pi}{3}\rho_q =  \frac{M}{R^3},\qquad \frac{\Theta_q}{3}=\frac{\dot R}{R},\end{equation}
 transforming \eqref{Flike3} into a relation between q-scalars that is formally identical to the Friedman equation of FLRW dust models:
\begin{equation}\left(\frac{\Theta_q}{3}\right)^2=\frac{8\pi}{3}\rho_q-\frac16\,{}^3{\cal R}_q,\end{equation}
with ${}^3{\cal R}_q$ given by \eqref{3Rq}.  

Notice that CG allows to incorporate into the spatial curvature a position dependent  term $\lambda(r)$ analogous to the cosmological constant. In fact, as in FLRW models, by setting $K=\gamma=0$ and  ${}^3{\cal R}=(8\pi/3)\Lambda=-\lambda(r)$ in \eqref{3RLTB}, a positive cosmological constant  can be characterized as the particular case of LTB models in CG with constant negative spatial curvature. 
 
As a consequence of the modification imposed by CG that results in the Friedman-like equation \eqref{Flike3}, the evolution of comoving dust layers of this CG solution in general is found by integrating the quadrature 
\begin{equation} t-t_{bb}(r) =\int_0^R{\frac{\sqrt{\bar R}\,d\bar R}{\sqrt{2M-K \bar R-\gamma\,\bar R^2-\lambda\,\bar R^3}}},\label{quadrature}\end{equation}
where  $t_{bb}(r)$ is an integration constant denoting the ``big bang time''. As opposed to  FLRW models for which CG  does not provide an evolution equation for the term ${\cal K}(a)$, in the case of LTB models of CG the field equations provide a fully determined form of ${\cal F}(R)$ in \eqref{eqPhiLTB}. Hence, the initial value formulation is well posed, with the evolution of the models fully determined at all times by initial conditions given by specifying at an initial hypersurface orthonormal to the 4-velocity the functions $K(r),\,M(r),\,t_{bb}(r),\,\gamma(r)$ and $\lambda(r)$ (though one of these functions can be eliminated by a choice of radial coordinate, for example $R_0=R(t_0,r)=r$). 

The presence of the term $\gamma R^2$ means that solutions of \eqref{quadrature} are manifestly distinct from those of the equivalent quadrature for the LTB dust solution of GR with $\Lambda$ zero or nonzero (the case $\gamma=0$). For the usual LTB models in GR ($\Lambda=0$) the spatially flat sub-case  ($K=0$) dust layers are ever expanding, while the equivalent CG solution ($K=0,\,\, \gamma\ne 0$) allows for collapsing layers if $\gamma> 0$. Another interesting feature is the radial dependence of $\lambda(r)$ analogous to the cosmological constant $\Lambda$. Hence LTB models can model inhomogeneities in which the effect of $\Lambda$ is negligible ($\lambda = 0$ near $r=0$) and becomes dominant ($\lambda \to 8\pi \Lambda/3$ as $r\to\infty$). Also, accelerated expansion of dust layers is possible (even with $\Lambda=0$) in many cases in which the layers decelerate in the GR solutions. Since LTB dust solutions are simple but useful models of structure formation, this difference of evolution due to modifications from CG might have interesting consequences.

If $M,\,\gamma,\,\lambda$ are constant the LTB solution \eqref{Flike3} from CG also reduces to the Schwarzschild (or Schwarzschild-Kottler if $\lambda\ne 0$)  CG solution \eqref{Schw2}, but described in comoving coordinates constructed with radial timelike geodesics. This limit provides a consistency test to the Codazzi formulation of CG, since it is parallel to the vacuum limit of LTB models in GR to the Schwarzschild solution (or Schwarzschild-Kottler if $\lambda\ne 0$) in the same comoving coordinates based on timelike radial geodesics.  

The FLRW limit of LTB dust models in CG is analogous to the same limit in GR: it follows as $R$ becomes a separable function  $R=a(t) f(r)$, with $a(t)$ satisfying the modified Friedman equation  \eqref{friedeqCT} with $\gamma {\cal K}=\gamma a +\lambda a^2$. While the linear term $\gamma a$ of ${\cal K}$ is not compatible with a well posed initial value formulation when deriving the FLRW models, in this case it would follow as a limit of LTB models complying with this formulation.  

Finally, the description of the LTB dust solution in CG in terms of an ``effective'' energy momentum tensor $T_{(\hbox{\tiny{mod}})}$, as described in \eqref{effective}, leads to an unphysical fluid with anisotropic pressure whose components are
\begin{eqnarray}
-T^t_t {}_{(\hbox{\tiny{mod}})}&=& \rho + c_3 + \frac{2c_1}{R},\label{TCLTB1}
\\ T^r_r{}_{(\hbox{\tiny{mod}})} &=& -c_3 - \frac{2c_1}{R},\label{TCLTB2}\\
T^\theta_\theta{}_{(\hbox{\tiny{mod}})} &=& T^\phi_\phi{}_{(\hbox{\tiny{mod}})}= - c_3- \frac{c_1}{R},\label{TCLTB3}
\end{eqnarray}
with $\rho$ given by \eqref{LTBdust}. The ``effective'' fluid satisfies $T^{ab}_{(\hbox{\tiny{mod}})}\,_{;b}=0$ as long as $c_1^\prime= c_3^\prime=0$. However, it is not compatible with a regular center worldline ($\rho,\,p$ bounded at $r=0$ with $R(t,0)=\dot R(t,0)=0$), hence it cannot describe a model of a localized spherical inhomogeneity. Also, this fluid has no FLRW limit as $R$ becomes a separable function  $R=a(t) f(r)$.

\section{Szekeres solutions} 

Szekeres models are a large class of exact solutions, typically with a comoving dust source ($T^{ab} =\rho u^a u^b,\,\,u^a=\delta^a_t$),  admitting (in general) no isometries. They are expressible in the line element
\begin{equation} ds^2 =-dt^2 + e^{2\alpha} dr^2 + e^{2\beta} [dx^2+dy^2]\label{sezk1}\end{equation}
where $\alpha,\,\beta$ depend on all four coordinates. We consider class I models  ($\beta' =\beta_{,r} \ne 0$) and the quasi-spherical sub-class of solutions whose spherically symmetric limit are LTB models. For this solution \eqref{sezk1} is expressible in the following LTB-like parametrization
\begin{eqnarray}. ds^2 &=& -dt^2 + \frac{{\cal E}^2 Y^\prime{}^2}{1-K}\,dr^2 + Y^2 [dx^2+dy^2], \label{szek2}\\
Y &=& \frac{R(t,r)}{{\cal E}(r,x,y)}, \quad K=K(r),
\end{eqnarray}
with ${\cal E}$ and $Y$ satisfying 
\begin{eqnarray} {\cal E} &=&  \frac{S}{2}\left[1+\left(\frac{x-P(r)}{S}\right)^2+\left(\frac{y-Q(r)}{S}\right)^2\right],\\
\dot Y^2 &=& \frac{2\tilde M}{Y} - \tilde K \quad \Rightarrow\quad \dot R^2 = \frac{2M}{R}-K, \label{friedeqSz}
\end{eqnarray}
with $ \tilde M= M(r)/{\cal E}^3$ and $\tilde K = K/{\cal E}^2$ and we assume $\Lambda=0$. Einstein's field equations reduce to the single nonzero component:
\begin{equation}\ G_{tt} = 8\pi \rho = \frac{2\tilde M^\prime}{Y^2\,Y^\prime}. \end{equation}
 In this parametrization the Friedman-like equation and the nonzero component of the field equations of the Szekeres solution are formally identical to their LTB equivalents  \eqref{Flike} and \eqref{LTBdust}. Therefore, the CG Szekeres solution follows the same pattern as the LTB solution in equations, with the Friedman-like equation
  \begin{eqnarray} \dot Y^2 = \frac{2\tilde M}{Y} - \tilde K - \tilde{\cal F},\qquad  \tilde{\cal F}=\frac{{\cal F}(R)}{{\cal E}^2}\nonumber\\
  \Rightarrow\quad \dot R^2 = \frac{2M}{R} - K - {\cal F}\end{eqnarray}
 leading to the same components of the prospective Codazzi tensor $ \mathscr{C}_{ab}$ in \eqref{codazzi3a} for the LTB metric  in \eqref{codLTB1}-\eqref{codLTB2}, with quantities normalized by ${\cal E}^2$ and the constraint \eqref{consLTB} by ${\cal E}^3$. All comments and results on degrees of freedom that CG affords to LTB models apply also to the much more general  quasi-spherical Szekeres models, which provide more elements for modeling cosmological inhomogeneities. 
 
 \section{A spherically symmetric static solution}

Consider the spherically symmetric line element for static GR solutions
\begin{equation}ds^2=-\exp[2\Phi(r)] dt^2+\frac{dr^2}{1-\frac{2 M(r)}{r}}+r^2\, (d\theta^2+\sin^2\theta d\phi^2),\label{static} \end{equation} 
where  the source is a perfect fluid $T^{ab}=\rho u^a u^b + ph^{ab}$ with $u^a = \exp(-\Phi) \delta^a_t$. Einstein's field equations and the conservation law $\nabla_bT^{ab}=0$
\begin{eqnarray}    
M^\prime  &=& 4\pi \rho r^2,\label{feqst1} \\
  \Phi^\prime &=&  \frac{4\pi\, p\, r^3+M}{r(r-2M)},\label{feqst2}\\
  p^\prime   &=& -(\rho+p)\Phi^\prime,\label{mombalst}
\end{eqnarray}
determine the static solution once an equation of state linking $p$ and $\rho$ is selected.
Using the equations above, the Codazzi tensor \eqref{codazzi} for GR
\begin{eqnarray}{\cal C}^r_r &=& {\cal C}^\theta_\theta={\cal C}^\phi_\phi = -\frac{8\pi}{3}\rho+\frac{2M^\prime}{3r^2}=0,\label{codGRstat1}\\
{\cal C}^t_t &=& 8\pi\left(\frac{2}{3}\rho+p\right)-2\left(1-\frac{2M}{r}\right)\frac{\Phi^\prime}{2} \nonumber\\
&+& \frac{2(3M-2rM^\prime)}{3r^3}=0.\label{codGRstat2}\end{eqnarray} 
so that  \eqref{feqst1}-\eqref{mombalst} inserted in \eqref{codGRstat1}-\eqref{codGRstat2} yield ${\cal C}^a_b=0$ and satisfies \eqref{codazziGR}.

The modification of rotation velocities of circular geodesics provides a motivation to obtain a nonzero Codazzi  tensor $ \mathscr{C}_{ab}$ in \eqref{codazzi3a} and thus a CG solution satisfying \eqref{codazzi3b}. From the particle Lagrangean $-2{\cal L}=-\exp(2\Phi)\dot t^2+r_0^2\dot \phi^2$, the velocity of observers along timelike circular geodesics ($\dot r =dr/d\tau=0, \theta=\pi/2$)  associated with \eqref{static}  are  
\begin{equation} v^2(r) = r^2\,\dot \phi^2 = r\left[e^{2\Phi(r)}-1\right]\approx r\Phi^\prime\approx \frac{M(r)}{r}, \label{rotvel}\end{equation}
where we used the weak field limit  $2M/r\ll 1,\,\,\Phi\ll 1$ also applied to  \eqref{feqst2} ($p/\rho \sim  \sigma^2\ll c^2$, with $\sigma^2$ the dispersion velocity).  Equation \eqref{rotvel} shows that in GR the only way to account for the observed increase of $v^2(r)$ in galactic rotation curves is an increase of $M$ (and of density $\rho$ from \eqref{feqst1}) and a modification of $\Phi^\prime$ produced by the assumption of an extra ``invisible'' mass.

Thus, CG, as an alternative gravity theory, can be useful to increase $v^2(r)$ without modifying the matter-energy source taken as visible mass and density  $M$ and $\rho$. For this purpose we need to pass from \eqref{codazziGR} to generate a nonzero Codazzi  tensor $ \mathscr{C}_{ab}$  in \eqref{codazzi3a} that satisfies \eqref{codazzi3b}. Since we keep the same $\rho,\,M$ and perfect fluid $T_{ab}$, the most convenient choice is to modify the hydrostatic equilibrium equation \eqref{feqst2} :
\begin{equation}  \Phi^\prime = -\Psi^\prime+  \frac{4\pi\, p\, r^3+M}{r(r-2M)},\label{feqstCG}\end{equation}
where extra term  $\Psi(r)$  becomes determined by inserting \eqref{feqstCG} into \eqref{codazzi3a}. The result is
\begin{equation}  \mathscr{C}^t_t = -2\left(1-\frac{2M}{r}\right)\frac{\Psi^\prime}{r},\quad  \mathscr{C}^r_r =  \mathscr{C}^\theta_\theta= \mathscr{C}^\phi_\phi =0,\end{equation}
which inserted into \eqref{codazzi3b} leads to the following quadrature and its first integral
\begin{eqnarray}.  \frac{\Psi^{\prime\prime}}{\Psi^\prime} &=&  -\Phi^\prime +\frac{2rM^\prime -4M+r}{r(r-2M)},\\
 \Psi^\prime &=& -\frac{C_0\,r^2\, e^{-\Phi}}{r-2M},\label{Psir}\end{eqnarray} 
where $C_0$ is a constant in $1/\hbox{length}^2$ units and we eliminated $\rho$ and $p$ from \eqref{feqst1} and \eqref{feqst2}. The field equation \eqref{feqst1} and conservation equation \eqref{mombalst} remain the same, but the equilibrium equation \eqref{feqstCG} is modified as
\begin{equation}  \Phi^\prime =\frac{(4\pi p+C_0\,e^{-\Phi})r^3+M}{r(r-2M)},\label{modeq}\end{equation}
so that the GR solution is recovered for $C_0=0$ (with \eqref{modeq} becoming \eqref{feqst2}). Notice that \eqref{modeq} does not involve $\rho$, hence it does not affect $M$ obtained from the field equation \eqref{feqst1} (can be the same $\rho$ and $M$ of GR). However, \eqref{modeq} and \eqref{mombalst} are coupled, so  $\Phi$ and $p$ are necessarily different from their GR forms. 

The perfect fluid solution obtained from solving the system  \eqref{feqst1}, \eqref{mombalst} and  \eqref{modeq} is valid up to the boundary $r=r_b$ of (say) a spherical galactic system, where the static solution should match with the equivalent Schwarzschild CG solution \eqref{Schw1}-\eqref{Schw2}. Harada assumed this matching in \cite{harada2022cotton}, but matching conditions in CG still need to be examined. 

The weak field limit can be characterized by $2M/r\ll 1,\,\,\Phi\ll 1$ and $4\pi\,p\, r^3 \ll M$ (since $4\pi\rho r^3\sim M$ and  $p/\rho \sim \sigma^2/c^2\ll 1$ with $\sigma^2$ a characteristic  velocity dispersion). Hence, \eqref{Psir} and \eqref{modeq} become 
\begin{equation} \Psi^\prime \approx -C_0\,r,\qquad \Phi^\prime \approx C_0\,r +\frac{M}{r^2},\label{weak1} \end{equation}
which shows in the context of a Newtonian limit the effect of the modification introduced by CG: an acceleration increasing linearly with $r$ (first term in $\Phi^\prime$ with $C_0$) together with the decaying Keplerian acceleration (second term).   

Velocities of circular geodesics follow from the Lagrangean associated with the potential $v^2/c^2\approx r\Phi^\prime$
\begin{eqnarray} \frac{v^2(r)}{c^2} \approx 
 r\Phi^\prime = C_0\,r^2+ \frac{M}{r}\label{rotvels} \end{eqnarray}
identifying the Keplerian term $M(r)/r$ that is dominant at small scales $r\approx 0$ and the CG contribution that grows and becomes dominant as $r$ increases.

To probe this approximation, we use the parameters of the Plummer sphere, a toy model of a spherical galaxy \cite{binney2011galactic}
\begin{equation}  \Phi_N(r) = - \frac{GM_0}{\sqrt{b^2+r^2}}\quad \frac{4\pi}{3}\rho(r)=\frac{ M_0 b^2}{\left(b^2+r^2\right)^{5/2}},\label{Plummer1} \end{equation}
where $ \Phi_N(r)$ is the purely Newtonian potential (\eqref{weak1} with $C_0=0$), $M_0,\,b$ are constants with mass and length units. We assume conventional values  $\rho_c=\rho(0)= 0.4\,\hbox{M}_\odot/\hbox{pc}^3$ and $v_e = 500\,$ km/s as the escape velocity \cite{daod2019density} and leave $C_0$ as an adjustable parameter that conveys the scale of CG. It is useful to  define a dimensionless radius and relate the remaining parameter $b$ as 
\begin{equation} x = \frac{r}{b},\quad  \frac{1}{b^2} = \frac{4\pi G\,\rho_c}{3v_e^2},\quad M_0=\frac{4\pi}{3}\rho_c b^3.\label{Plummer2}\end{equation}
For a spherical galaxy of total mass $M_0 = 10^{11}\hbox{M}_\odot$ and radius $r_b=20$\,kpc, we get $b=3.34$ kpc.  Using these parameters and choosing $1/\sqrt{C_0}=100$ kpc, we plot the rotation curves $v(r)$ normalized with $v_e$ (in km/s) from \eqref{rotvels} in figure 1, separating the purely Keplerian curve (blue), the CG contribution (red) and their sum (black). Notice that the CG contribution compensates the decay of the Keplerian curve, with the sum of both producing a roughly flat profile as $r$ increases up to $r_b =20$ kpc. 

The increase of rotation velocities caused by CG occurs without modifying the GR density and mass in each spherical layer (as determined by the mass-luminosity relation).  While rotation velocities grow too fast $v\sim r$ as $r$ increases, the radial range is bounded by $0\leq r\leq r_b$, the boundary radius at which the fluid CG solution should match the Schwarzschild analogue \eqref{Schw2}. 

Evidently, this toy model is unrealistic,  though it serves to illustrate the possibility that extra degrees of freedom of CG can be useful to address the rotation velocities of galactic systems, either without assuming dark matter or assuming lesser quantity. Harada has already proven that CG can fit galactic rotation velocities without dark matter through the Schwarzschils analogue, but he did not actually derive the CG solution in the inner fluid region.  
\begin{figure}
\centering
  \includegraphics[width=1\linewidth]{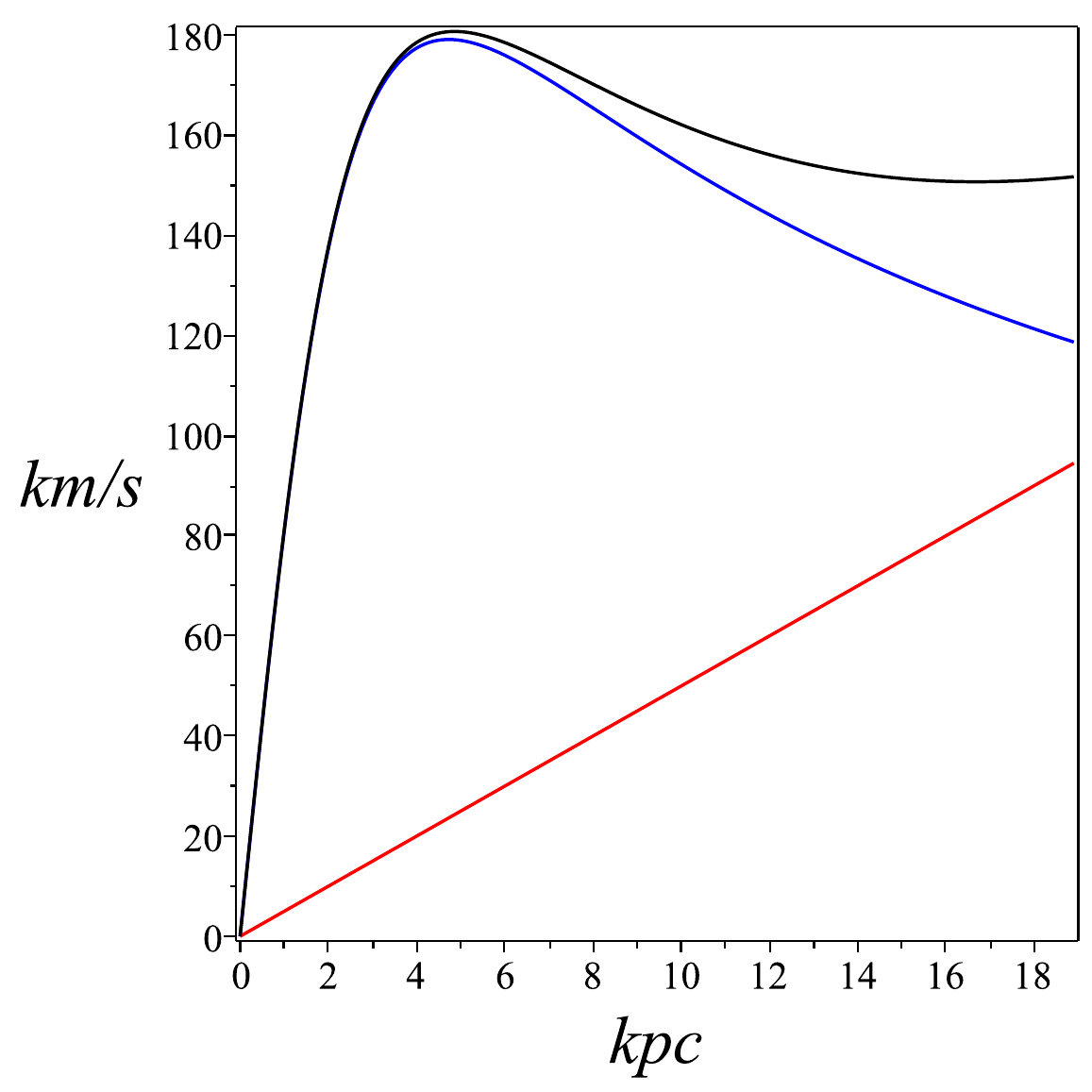}
 \caption{Rotational curves for a Plummer sphere. The curves respectively correspond to the Keplerian contribution (blue), the CG contribution (red) and their sum (black). They were obtained from \eqref{rotvels} with the parameters defined in \eqref{Plummer1}-\eqref{Plummer2}. Notice how the CG contribution lifts and flattens the decaying Keplerian curve that would result in the weak field of GR without sark matter. }
\end{figure} 

The matter-energy source of the CG solution we discuss in this section is $T^{ab}$, but it can also be expressed in terms of an ``effective'' energy momentum tensor $T_{ab}^{(\hbox{\tiny{eff}})}$, as described in \eqref{effective}, leading to a perfect fluid $T^{ab} = \rho u^a u^b + p^{(\hbox{\tiny{eff}})} h^{ab}$, with the modified pressure given by
\begin{equation} 8\pi p^{(\hbox{\tiny{eff}})} = 8\pi p-2\left(1-\frac{2M}{r}\right)\frac{\Psi^\prime}{r}.\label{pcot}\end{equation}
This modified pressure also follows from solving .\eqref{modeq} and \eqref{mombalst} with $\Psi'$ given by \eqref{Psir}, though $\rho$ and $M$ are not modified. 

\section{Spherically symmetric shear-free solutions}

A large class of solutions describing inhomogeneous and anisotropic  perfect fluid sources $T^{ab}=\rho u^a u^b+p h^{ab}$ with a shear-free 4-velocity ($\sigma_{ab}=0$, the spherically symmetric Stephani-Barnes solutions in \cite{Krasinski}) is given by the metric
 \begin{equation} ds^2 = - N^2 dt^2 + B^2\left[dr^2 + r^2(d\theta^2+\sin^2\theta d\phi^2)\right],\label{sfmetric1} \end{equation}
 where $N$ and $B$ depend on $(t,r)$ and the comoving 4-velocity is $u^a = N^{-1} \delta^a_t$. The constraint of Einstein equations $h_a^c\nabla_b\sigma^b_c-h_a^b\nabla_b\Theta=0$ implies $\Theta=\Theta(t)$. The FLRW limit of the solutions follow from $B=a(t)$ and $N=1$, hence the time coordinate can be fixed as 
 \begin{equation}  N = \frac{B_{,t}/B}{\Theta/3},\qquad \frac{\Theta}{3}=\frac{S_{,t}}{S}, \label{sfmetric2} \end{equation}
 where $S(t)$ is an arbitrary function  analogous to the FLRW scale factor.  
 The condition of pressure isotropy  $G^r_r-G^\theta_\theta=0$ leads to the constraint 
 \begin{equation}   L_{yy} = F(y) L^2,\qquad L = \frac{1}{B},\qquad  y = \frac{r^2}{2}, \end{equation}
where $F(y)$ is related to the conformal invariant $\Psi_2$ that determines the Weyl tensor for Petrov type D spacetimes \cite{sussman1987spherically} and has only an electric part: $C_{ab}^{cd}=4(h_{[a}^{[c}+u_{]a}u^{[c}) E_{b]}^{d]}$ \cite{ellis2012relativistic}. The components of the electric Weyl tensor and the radial derivative of the density are given by 
\begin{eqnarray} E^r_r &=&  \frac23\,y\,L^3\,F = -2 E^\theta_\theta = -2 E^\phi_\phi, \\
8\pi \rho_{,y} &=& (2y F_{,y}+5F)\,L^3,\end{eqnarray}
Individual solutions follow by choosing the functional form of $F(y)$ and will produce two arbitrary functions of time ($S(t)$ and another one) as integration constants. In particular, the spherically symmetric conformally flat Stephani solution is case with $L_{yy}=F=0$ (so that $C_{ab}^{cd}=0$ and $\rho=\rho(t)$) , while the McVittie solution corresponds to $F(y)=y^{-5/2}$.  
 
The conservation equations $\nabla_bT^{ab}=0$ are
 \begin{eqnarray}\rho_{,t} &=& -\frac{3L_{,t}}{L}(\rho+p),\quad h_a^b p_{,b}=-\dot u_a (\rho+p),\label{Tabdiv}\\
  \dot u_a &=& h_a^b \frac{N_{,b}}{N}=\left(\frac{L_{,yt}}{L_{,t}}-\frac{L_{,y}}{L}\right)\delta_a^y,\label{accSF}\end{eqnarray}
 while the GR density and pressure are given by
 \begin{eqnarray}\frac{8\pi}{3} \rho &=& \frac{S_{,t}^2}{S^2}+\left(\frac23 L^2F-L_{,y}^2\right)y +LL_{,y},\label{shfr1}\\
 p &=&-\rho+\frac{3L}{L_{,t}}\rho_{,t},\label{shfr2}\end{eqnarray}
 where \eqref{shfr2} follows from \eqref{Tabdiv}. Inserting the GR equations in \eqref{codazzi} yields ${\cal C}^a_b=0$, complying with \eqref{codazziGR} . To pass from \eqref{codazziGR} to \eqref{codazzi3a} to obtain a nonzero Codazzi tensor $\mathscr{C}_{ab}$ we modify ${\cal C}^a_b$ adding to ${\cal C}^i_i$ a function $\alpha(t,y)$ and to ${\cal C}^t_t$ a function $\beta(t,y)$. Substitution of  $\mathscr{C}_{ab}\ne 0$   into \eqref{codazzi3b}  yields three constraints whose solution is
\begin{equation} \alpha_{,y}=0,\qquad \beta = \alpha(t)+ \alpha_{,t}\frac{L}{L_{,t}},\label{ab} \end{equation}
which suggests identifying $\alpha$ with a modification of spatial curvature following the steps of \eqref{KCG2} (since $\Theta=\Theta(t)$), while the function $\beta$ can be identified with a modification of $L/L_{,t}$ that appears in the lapse function $N$ and alters the relation between proper and coordinate time: $\tau =\int{dt/N}$. In terms of an ``effective'' fluid as in \eqref{effective}, the functions $\alpha,\,\beta$ involve a modified effective energy momentum tensor  $T_{ab}^{(\hbox{\tiny{eff}})}$, with  $\rho_{(\hbox{\tiny{eff}})}=\rho-3\alpha$ and $p_{(\hbox{\tiny{eff}})}=p-2\alpha-\beta$.  This corroborates the result of  \cite{mantica2023codazzi}  that these solutions admit an ``effective'' energy momentum tensor of the form of a perfect fluid.
 
 The conformally flat case (Stephani solution $L_{,yy}=F=0$) is particularly interesting, as it is impossible to consider it in the original Cotton CG formulation of Harada (equations \eqref{original}-\eqref{cotton}), since the Cotton tensor vanish automatically. However, there is no problem in considering it in the formulation we used based on \cite{mantica2023codazzi}. For this case the GR solution from \eqref{sfmetric2}  and \eqref{shfr1}-\eqref{shfr2} lead to $\rho=\rho(t)$ and a Friedman-like equation analogous to that of FLRW models (though $p=p(t,y)$ through $1/N$).
 \begin{eqnarray} B &=& \frac{1}{L}= \frac{S}{1+K y},\label{Bfunct}\\
  N &=& -\frac{L_{,t}/L}{S_{,t}/S}= \frac{1+(K-K_{,t} S/S_{,t})y}{1+Ky},\label{Nfunct}\\
\frac{S_{,t}^2}{S^2} &=&  \frac{8\pi}{3}\rho(t)-\frac{K}{S^2},\label{FriedST}\\
8\pi p  &=& -\frac{S_{,t}^2}{S^2}-\frac{K}{S^2}\frac{2SS_{,tt}-2S_{,t}^2-K_{,t} +2K}{N}\label{pfunct}.
 \end{eqnarray}
The Hamiltonian constraint \eqref{KCG2} for the Stephani solution is similar to that of FLRW: ${}^3{\cal R} = 6K/S^2$, hence \eqref{FriedST} suggest that the function $\alpha(t)$ in \eqref{ab} can be a modification spatial curvature, while $\beta$ involves a modification of $1/N$ in  \eqref{pfunct}.  However, the evolution of the models might not follow a well posed initial value problem, since there are no constraints on the function $K(t)$, but this also occurs in the GR solutions. 

\section{Conclusion}

We have  presented an intuitive procedure for using the Codazzi formulation  to obtain non-trivial exact solutions in ``Cotton Gravity'' CG, a gravity theory alternative to General Relativity (GR), originally proposed by Harada \cite{harada2021emergence} in terms of the Cotton tensor, but (as we show in this article) more appropriately described by  the Codazzi formulation derived by Mantica and Molinari in \cite{mantica2023codazzi}. 

The Codazzi formulation corrects various ambiguities and inconsistencies of the Cotton formulation. Fore example, it allows to apply CG to non-vacuum conformally flat spacetimes (an impossible task in the original formulation because the Cotton tensor vanishes identically). The field equations of the Codazzi formulation directly involve the Einstein (or Ricci) and energy momentum tensors, removing ambiguities on the field sources and facilitating a natural correspondence to GR, an important requirement in alternative gravity theories. 

These advantages facilitates extend and generalize known exact solutions of GR as self-consistent CG solutions, with their same energy momentum tensors. We described a simple procedure for generating CG solutions from GR solutions, On vacuum solutions in static spherical symmetry, we re-derive the Schwarzschild analogue found in \cite{harada2021emergence} and provided arguments that point out to its unicity, though we still have no rigorous proof that CG complies with Birkhoff's theorem. We also obtained the Reissner-Nordstrom analogue and CG generalizations of well known solutions of GR: FLRW, LTB and Szekeres dust models, as well as perfect fluid spherically symmetric solutions: static and non-static with a shear-free 4-velocity. 

We found that CG brings an interesting geometric degree of freedom in FLRW, LTB and Szekeres models: it modifies the spatial curvature in the Hamiltonian constraint, without modifying the energy momentum tensor ({\it i.e.} without adding extra sources). In FLRW models this modification of spatial curvature must be necessarily restricted to describe the evolution as a well posed initial value problem, leading to FLRW models that are operationally the same as GR models, but with an important conceptual difference: the cosmological constant now interpreted as spatial curvature. In particular, the $\Lambda$CDM model becomes defined in a precise covariant manner as the unique FLRW dust model in CG with constant negative curvature, an appealing theoretical interpretation of the $\Lambda$ term. 

In LTB and Szekeres dust models the modification of spatial curvature is compatible with a well posed initial value problem, leading to manifestly distinct dynamical evolution of the models with respect to their GR equivalents. In particular, the LTB and Szekeres solutions in CG allow to describe dynamical effects absent in GR, such as a spatial and temporal dependent transition from a decelerated expansion to an accelerated expansion of dust layers, an evolution produced only by dust sources and  spatial curvature, without assuming the presence of dark energy fluids with negative pressure or imposing a cosmological constant. 

In static fluid spheres, we show that CG qualitatively modifies the radial profile of rotation curves of circular geodesics. Considering as source only visible matter, without assuming dark matter, these curves can mimic in the weak field limit  the flattening behavior with increasing radius that is associated with galactic dark matter. We show this effect in a toy model of a spherical galaxy treated in the weak field limit. Finally, we apply CG to spherically symmetric solutions with a shear-free 4-velocity, corroborating the results of \cite{mantica2023codazzi}.

We believe that CG has a significant potential as an alternative gravity theory, as it is closer to GR (and less technically complicated) in comparison with other alternative gravity theories. The procedure to obtain CG solutions is the first step towards a more rigorous and solutions we have presented are highly idealized, but out aim has been to show how  CG solutions can be generated. In future research we will attempt to find more general models.  

\appendix

\section{Equivalence between the Cotton and Codazzi formulations CG}

We show in this Appendix that the Cotton and Codazzi formulations of CG lead to the same field equations. However, as we have explained in this article, the way these equations are understood and solved is considerably different, with the Codazzi formulation providing deeper theoretical insight and ease to find solutions.      

The CG field equations in the Cotton formulation given by \eqref{original}-\eqref{Tcot} using the Einstein tensor instead of the Ricci tensor by the substitution ${\cal R} = -{\cal G},\,\,{\cal R}_{ab}=G_{ab}-\frac12 G\,g_{ab}$:
\begin{eqnarray}C_{abc} &=& 16\pi {\cal T}_{abc},\label{E1},\\
C_{abc} &=& \nabla_b G_{ac}-\nabla_c G_{ab}-\frac13\left(g_{ac}\nabla_bG-g_{ab}\nabla_cG \right),\nonumber\\\label{E2} \\
{\cal T}_{abc} & = & \nabla_{a}T_{bc}-\nabla_{b}T_{ac}-\frac{1}{3}\left(g_{bc}\nabla_{a}T-g_{ac}\nabla_{b}T\right),\nonumber\\\label{E3}
\end{eqnarray}
where $C_{abc}$ is the Cotton tensor \cite{cotton1899varietes, stephani2009exact,garcia2004cotton} and ${\cal T}_{abc}$ is related to the generalized angular momentum tensor. Equations \eqref{E2} and \eqref{E3} can be rewritten as 
\begin{eqnarray}
C_{abc}=\nabla_b{\cal S}_{ac}-\nabla_c{\cal S}_{ab},\quad 
{\cal T}_{abc} = \nabla_b\mathscr{T}_{ac}-\nabla_c\mathscr{T}_{ab},\nonumber\\\label{E4}
\end{eqnarray}
where the ${\cal S}_{ab}$ is the Schouten tensor in \eqref{schouten} in terms of $G_{ab}$, with $\mathscr{T}_{ab}$ given by
\begin{eqnarray}  {\cal S}_{ab} =  G_{ab}-\frac13 \,G g_{ab},\label{E5}\quad
\mathscr{T}_{ab} =T_{ab}-\frac13\,T\,g_{ab}, \label{E6} \end{eqnarray}
Using \eqref{E4}-\eqref{E6} the Cotton field equations \eqref{E1} become
\begin{eqnarray}
\nabla_b\left({\cal S}_{ac}-8\pi \mathscr{T}_{ac}\right) - \nabla_c\left({\cal S}_{ab}-8\pi \mathscr{T}_{ab}\right) =0,
\end{eqnarray}
which identically coincide with the field equations in the Codazzi formulation of Mantica and Molinari \cite{mantica2023codazzi} (equation \eqref{codazzi2})
\begin{eqnarray}{\cal C}_{ab} &=& {\cal S}_{ab}-8\pi\mathscr{T}_{ab} \quad  \hbox{ is a Codazzi tensor }\nonumber\\
 &\Rightarrow&\quad  \nabla_bC_{ac}-\nabla_cC_{ab}=0,\label{E7}\end{eqnarray}
The reverse equivalence (from \eqref{E7} to \eqref{E1}) follows by the same steps in reverse.

\section{Ambiguities in sphericaly symmetric vacuum solutions in the Cotton formulation}

The field equations in the Cotton formalism are \eqref{original} $C_{abc}=8\pi {\cal T}_{abc}$, where $C_{abc}$ is the Cotton tensor and $ {\cal T}_{abc}$  is the generalized angular momentum given by \eqref{Tcot}. These equations suggest, that finding a vacuum solution in static spherical symmetry  requires setting $ {\cal T}_{abc}=0$ and solving for $C_{abc}=0$ for the metric \eqref{staticSS}. This is quite problematic: while $T_{ab}=0$ implies $ {\cal T}_{abc}=0$, we prove in this Appendix that $ {\cal T}_{abc}=0$ does not imply $T_{ab}=0$.  

The condition $C_{abc}=0$ for the metric \eqref{staticSS} leads to the following complicated third order equation
\begin{eqnarray}
\frac{A'''}{A}+\left(\frac{3B'}{2B}-\frac{2A'}{A}+\frac{2}{r}\right)\frac{A''}{A}+\left(\frac{A'}{A}-\frac{B'}{B}-\frac{1}{2r}\right)\frac{A'^{2}}{A^{2}}\nonumber\\
+\left(\frac{B''}{2B}+\frac{B'}{2rB}-\frac{2}{r^{2}}\right)\frac{A'}{A}+\frac{2}{r^{2}}\left(1-\frac{1}{B}\right)=0.\nonumber \\
\label{B0}
\end{eqnarray}
which admits an infinite number of solutions, though it reduces to \eqref{ODE3} if setting $B=A$.

However, these solutions are not vacuum solutions. This can be proved using the most general energy momentum tensor compatible with \eqref{staticSS} (a fluid with anisotropic pressure) and its conservation equation $\nabla_{b}T^{ab}=0$
\begin{eqnarray}
T_{ab} & = & \rho u_{a}u_{b}+ph_{ab}+\Pi_{ab},\label{B1}\\
 (p-2P)' &=& -\frac{\rho+p-2P}{2}\frac{A'}{A}+\frac{3P}{r}, \label{B2}
\end{eqnarray}
where $\Pi_{ab}$ is the anisotropic pressure whose nonzero components are $\Pi_{r}^{r}=-2P,\,\,\Pi_{\theta}^{\theta}=\Pi_{\phi}^{\phi}=P,\,\,\Pi_{t}^{t}=0$, with $P=P(r)$. Substitution of this energy momentum tensor in the definition of ${\cal T}_{abc}$  in \eqref{Tcot} leads to 
\begin{equation}
p'=-\frac{[\rho+p-2P]\,A'}{2A}-\frac{2}{3}\rho',\label{B3}
\end{equation}
which combined with \eqref{B2} leads to
\begin{equation}
P'=-\frac{1}{3}\rho'-\frac{3P}{r}.\label{B4}
\end{equation}
As a consequence, condition ${\cal T}_{abc}=0$ does not determine in general a vacuum solution. It can correspond to a vacuum solution if $\rho=p=P=0$, but \eqref{B0} admits an infinite number of solutions characterized by fluids complying with \eqref{B1}-\eqref{B4}. However, for any given solution it is impossible to determine to which specific form of \eqref{B1} they correspond to. This ambiguity is very problematic for the Cotton formulation, but it does not arise in the Codazzi formulation. 


\begin{acknowledgments}.
SN acknowledges financial support from SEP–CONACYT postgraduate grants program
\end{acknowledgments}


\bibliographystyle{unsrt}
\bibliography{CodazziGravity3b}
\end{document}